\documentclass{article}
\usepackage{amsfonts}
\usepackage{amssymb}
\usepackage{graphicx}
\usepackage{amsmath}

\begin{document}

\title{Tomography in abstract Hilbert spaces}
\author{V.I. Man'ko \\
{\footnotesize \textit{P.N.Lebedev Physical Institute, Leninskii
Prospect
53, Moscow 119991, Russia }}\\
[2ex]G. Marmo, A. Simoni, F. Ventriglia \\
\textsl{\footnotesize{Dipartimento di Scienze Fisiche dell'
Universit\`{a} ``Federico
II" e INFN, Sezione di Napoli}}\\
\textsl{\footnotesize{Complesso Universitario di Monte S. Angelo,
via Cintia, 80126 Napoli, Italy}}
\\\textsl{\footnotesize {(e-mail: \texttt{manko@na.infn.it, marmo@na.infn.it, simoni@na.infn.it, ventriglia@na.infn.it})}}}
\maketitle

\begin{abstract}
The tomographic description of a quantum state is formulated in an abstract
infinite dimensional Hilbert space framework, the space of the
Hilbert-Schmidt linear operators, with trace formula as scalar product.
Resolutions of the unity, written in terms of over-complete sets of rank-one
projectors and of associated Gram-Schmidt operators taking into account
their non-orthogonality, are then used to reconstruct a quantum state from
its tomograms. Examples of well known tomographic descriptions illustrate
the exposed theory.
\end{abstract}

\section{Introduction}

Standard description of a quantum state is done by means of a vector in an
abstract Hilbert space \cite{Dirac}. There are also descriptions on phase
space of quasi-distributions like Wigner function \cite{Wig32}, Husimi-Kano
Q-function \cite{Hus40,Kano56}, Sudarshan-Glauber P-function \cite
{Sud63,Glau63}, all of them are used to represent quantum states, both pure
and mixed ones. Recently the optical probability distribution of homodyne
quadrature has been introduced \cite{Ber-Ber,Vog-Ris} as an approach to
reconstruct the Wigner function of a quantum state.

This approach has been extended and symplectic tomography of a quantum state
has been suggested \cite{Mancini95} for reconstructing the Wigner function.
The description of a quantum states by means of a tomographic probability
distribution (tomogram) has been used to suggest a new formulation of
quantum mechanics \cite{Mancini97PL-Found Phys} in which these probability
distributions identify quantum states alternatively to vectors in a Hilbert
space, or density operators for mixed states. Due to the development of
tomographic methods, it turns out that there exists a map of the elements of
a Hilbert space (vectors) onto elements of the set of the probability
distributions (tomograms of a quantum state). The general problem how to
construct the tomographic map and what is the explicit mathematical
mechanism providing this map in abstract Hilbert spaces has been studied in
\cite{PLA2} for the finite dimensional case. The main idea of this mechanism
is to consider the squared modulus of the scalar product of two vectors in
the initial abstract Hilbert space $\mathcal{H}$ as the standard scalar
product of other vectors in another Hilbert space $\mathbb{H},$ which is the
space of operators acting on the initial space $\mathcal{H}$ As it has been
established in Ref. \cite{PLA2}, the approach which uses two Hilbert spaces
and the completeness condition, related to any (over-) complete set of
rank-one projectors in the new Hilbert space $\mathbb{H},$ naturally
provides the necessary ingredients to construct the tomographic map under
discussion.

The aim of the present work is to review the mathematical mechanism of the
tomographic map in finite-dimensional Hilbert spaces and to extend the
construction to the case of infinite dimensional Hilbert spaces. We will
clarify how the known examples of tomographic maps in the infinite
dimensional Hilbert spaces, like symplectic tomography \cite{d'Ariano96} and
photon number tomography \cite{BanWog,WogW,TombesEPL}, correspond to our
formulation for constructing a tomographic map in an abstract Hilbert space.
Moreover we reconsider briefly the known non-negative quasi-distribution
which is the Husimi-Kano $Q-$function from the point of view of
\textquotedblleft coherent state tomography\textquotedblright . As it is
shown in Ref. \cite{PLA3}, the construction of the Husimi-Kano $Q-$function
can be interpreted as finding a specific tomographic set of basis vectors
which provides the possibility to reconstruct the density operator in terms
of the $Q-$function exactly in the framework of our tomographic approach in
an abstract Hilbert space.

The paper is organized as follows. In the next section 2 the tomographic map
in finite-dimensional abstract Hilbert space is reviewed. In section 3 the
tomographic sets in Hilbert spaces are discussed in a general setting. In
section 4 minimal tomographic sets and resolutions of identity are
considered in infinite dimensional Hilbert spaces. Problem of operators
generating tomographic sets and examples of squeeze tomography and
symplectic tomography as well as a finite dimensional counter-example are
considered in section 5. \textit{Skewness} (i.e., completeness) of basis
vectors, coherent state and photon number tomographies are discussed in
section 6. Some conclusions and perspectives are drawn in section 7.

\section{The finite dimensional case}

In a previous work \cite{PLA2}, we have provided an interpretation of
quantum tomography in an abstract, finite-dimensional, Hilbert space $%
\mathcal{H}$ in terms of complete sets of rank-one projectors $\left\{
P_{\mu }\right\} _{\mu \in M}$, where $M$ is a set of (multi-) parameters,
discrete or continuous, collectively denoted by $\mu $. In general, a
tomogram of a quantum state $\left| \psi \right\rangle $ is a positive real
number $\mathcal{T}_{\psi }(\mu )$ depending on the parameter $\mu $ which
labels a set of states $\left| \mu \right\rangle \in \mathcal{H},$ defined
as
\begin{equation}
\mathcal{T}_{\psi }(\mu ):=\left| \left\langle \mu |\psi \right\rangle
\right| ^{2}.
\end{equation}
Our main idea was to regard the tomogram $\mathcal{T}_{\psi }(\mu )$ as a
scalar product on the (Hilbert) space $\mathbb{H}$ of the rank-one
projectors $\left| \mu \right\rangle \left\langle \mu \right| =P_{\mu
}\rightarrow \left| P_{\mu }\right\rangle \in \mathbb{H}:$%
\begin{equation}
\mathcal{T}_{\psi }(\mu )=\mathrm{Tr}\left( P_{\mu }\rho _{\psi }\right)
=:\left\langle P_{\mu }|\rho _{\psi }\right\rangle .  \label{Tomo}
\end{equation}
Equation (\ref{Tomo}) may readily be used to define the tomogram of any
density operator $\hat{\rho}$ or any other (bounded) operator $\hat{A}$%
\begin{equation}
\mathcal{T}_{A}(\mu ):=\mathrm{Tr}\left( P_{\mu }\hat{A}\right)
=\left\langle P_{\mu }|A\right\rangle .  \label{Tomogen}
\end{equation}
Equation (\ref{Tomogen}) shows in general that the tomogram of the operator $%
\hat{A}$ may be thought of as a \textit{symbol} of $\hat{A}:$ in other
words, by means of the set $\left\{ \left| \mu \right\rangle \left\langle
\mu \right| \right\} _{\mu \in M}$, to any operator $\hat{A}$ a function $%
\left\langle \mu |\hat{A}|\mu \right\rangle $ of the variables collectively
denoted by $\mu $ corresponds in a given functional space. For instance, in
the case of the symplectic tomography, the variables $\mu $ vary in the
phase space $M$ of the physical system. So, a tomography may be thought of
as a de-quantization, and in fact we found useful to study the
quantum-classical transition by comparing classical limits of quantum
tomograms with the corresponding classical tomograms \cite{PLA1}. Of course,
while the correspondence $\hat{A}\rightarrow \mathcal{T}_{A}(\mu )$ may be
thought of as a de-quantization, the inverse correspondence $\mathcal{T}%
_{A}(\mu )\rightarrow \hat{A}$ may be considered to give a quantization. The
symbol determines completely the operator: the reconstruction of the
operator $\hat{A}$ from its tomogram $\mathcal{T}_{A}(\mu )$ may be written
as:
\begin{equation}
\hat{A}=\sum\limits_{\mu \in M}\hat{K}_{\mu }\mathrm{Tr}\left( P_{\mu }\hat{A%
}\right) \Leftrightarrow \left| A\right\rangle =\sum\limits_{\mu \in
M}\left| K_{\mu }\right\rangle \left\langle P_{\mu }|A\right\rangle .
\end{equation}
In other words, the reconstruction of any operator is possible because the
tomographic set $\left\{ P_{\mu }\right\} _{\mu \in M}$ provides a
resolution of the identity (super-) operator\footnote{%
In the present paper we do not address the problem of the continuity of the
reconstruction formula \cite{Grossmann}, which is granted in all our
examples.} on $\mathbb{H}$%
\begin{equation}
\mathbb{\hat{I}}=\sum\limits_{\mu \in M}\hat{K}_{\mu }\mathrm{Tr}\left(
P_{\mu }\cdot \right) =\sum\limits_{\mu \in M}\left| K_{\mu }\right\rangle
\left\langle P_{\mu }\right| .
\end{equation}
Here the $\hat{K}_{\mu }$'s are Gram-Schmidt operators, which take into
account that in general the projectors $P_{\mu }$'s are not orthogonal,
while the sum may be an integral with a suitable measure. Thus, for the
finite $n-$dimensional case, $\mathbb{H=}\mathcal{H\otimes H}$ is $n^{2}-$%
dimensional and a \textit{minimal} tomographic set is a basis $\left\{
P_{k}\right\} $, ${k\in \left\{ 1,...,n^{2}\right\} },$ of rank-one
projectors which may be orthonormalized by a Gram-Schmidt procedure
\begin{equation}
\left| V_{j}\right\rangle =\sum\limits_{k=1}^{n^{2}}\gamma _{jk}\left|
P_{k}\right\rangle \quad ,\quad \left\langle V_{i}|V_{j}\right\rangle
=\delta _{ij}.
\end{equation}
In general, every element of the orthonormal basis $\{\left|
V_{j}\right\rangle \}$ is a linear combination of projectors, rather than a
single projector like $\left| P_{k}\right\rangle .$ Then a resolution of the
unity on $\mathbb{H}$ in terms of the $P_{k}$'s reads as
\begin{equation}
\mathbb{\hat{I}}_{n^{2}}=\sum\limits_{i=1}^{n^{2}}\left| V_{i}\right\rangle
\left\langle V_{i}\right| =\sum\limits_{i,j,l=1}^{n^{2}}\gamma _{il}^{\ast
}\gamma _{ij}P_{j}\mathrm{Tr}(\hat{P}_{l}\cdot
)=\sum\limits_{l=1}^{n^{2}}\left| K_{l}\right\rangle \left\langle
P_{l}\right| =\sum\limits_{j=1}^{n^{2}}\left| P_{j}\right\rangle
\left\langle K_{j}\right|
\end{equation}
where the Gram-Schmidt operator $\hat{K}_{l}$ has been introduced
\begin{equation}
\left| K_{l}\right\rangle =\sum\limits_{i=1}^{n^{2}}\gamma _{il}^{\ast
}\left| V_{i}\right\rangle =\sum\limits_{i,j=1}^{n^{2}}\gamma _{il}^{\ast
}\gamma _{ij}\left| P_{j}\right\rangle .
\end{equation}
We observe that $\hat{K}_{l}$ is a nonlinear function of the projectors $%
P_{k}$, because also the coefficients $\gamma $'s depend on the projectors.
Moreover, it results
\begin{equation}
\left\langle P_{i}|K_{l}\right\rangle =\sum\limits_{j=1}^{n^{2}}\gamma
_{jl}^{\ast }\left\langle P_{i}|V_{j}\right\rangle
=\sum\limits_{j,k=1}^{n^{2}}\gamma _{jl}^{\ast }(\gamma ^{\ast
})_{ik}^{-1}\left\langle V_{k}|V_{j}\right\rangle
=\sum\limits_{j=1}^{n^{2}}\gamma _{jl}^{\ast }(\gamma ^{\ast
})_{ij}^{-1}=\delta _{il}.  \label{OrtoPK}
\end{equation}

Similar formulae hold even for any other tomographic, i.e. (over-) complete,
set $\left\{ P_{\mu }\right\} _{\mu \in M}$ . For instance for the spin
tomography, in the \textit{maximal} qu-bit case $M=S^{2}$ is the Bloch
sphere of all rank-one projectors and we have \cite{PLA2}:
\begin{equation}
\mathbb{\hat{I}}=\int_{0}^{2\pi }\int_{0}^{\pi }\left| K(\theta ,\phi
)\right\rangle \mathrm{Tr}(P(\theta ,\phi )\cdot )\sin \theta d\theta d\phi ,
\end{equation}
where, in matrix form,
\begin{equation}
P(\theta ,\phi )=\frac{1}{2}\left[ \mathbb{I+}\vec{n}\cdot \vec{\sigma }%
\right] =\frac{1}{2}\left[
\begin{array}{cc}
1+\cos \theta & e^{-i\phi }\sin \theta \\
e^{i\phi }\sin \theta & 1-\cos \theta
\end{array}
\right]
\end{equation}
and
\begin{equation}
\hat{K}(\theta ,\phi )=\frac{1}{4\pi }\left[
\begin{array}{cc}
1+3\cos \theta & 3e^{-i\phi }\sin \theta \\
3e^{i\phi }\sin \theta & 1-3\cos \theta
\end{array}
\right] ,
\end{equation}
so that, for any operator $\hat{A},$ it results
\begin{eqnarray}
\hat{A} &=&\int_{0}^{2\pi }\int_{0}^{\pi }\hat{K}(\theta ,\phi )\mathrm{Tr}%
(P(\theta ,\phi )A)\sin \theta d\theta d\phi  \notag \\
&=&\int_{0}^{2\pi }\int_{0}^{\pi }P(\theta ,\phi )\mathrm{Tr}(\hat{K}(\theta
,\phi )A)\sin \theta d\theta d\phi .
\end{eqnarray}
This example shows that the orthogonality relations of the minimal case, Eq.
(\ref{OrtoPK}), do not hold in general.

After this brief introductory sketch of our previous work, we are ready to
extend our interpretation of tomography in abstract, infinite dimensional
Hilbert spaces

\section{Tomographic sets in Hilbert spaces}

Let $M$ be a set of (multi-) parameters $\mu $, and assign a map
\begin{equation}
\mu \in M\longrightarrow P_{\mu }\in \mathbb{P\subset H}
\end{equation}
from $M$ into the set $\mathbb{P}$ of all the rank-one projectors of the
Hilbert space $\mathbb{H}$. By definition, the set $\left\{ P_{\mu }\right\}
_{\mu \in M}$ is \textit{tomographic} if it is complete in $\mathbb{H}$. A
tomographic set determines a \textit{tomography} which is a functional,
linear in the second argument
\begin{equation}
\mathcal{T}:\mathbb{P}\times \mathbb{H}\longrightarrow \mathbb{C},(P_{\mu
},A)\longrightarrow \mathcal{T}_{A}(\mu )=\mathrm{Tr}\left( P_{\mu }\hat{A}%
\right) =\left\langle P_{\mu }|A\right\rangle .  \label{Tomography}
\end{equation}
This definition is appropriate in the finite $n-$dimensional case, where
\begin{equation}
\left| \mu \right\rangle \in \mathcal{H}_{n}\Leftrightarrow P_{\mu }\in
\mathbb{H}_{n^{2}}=B(\mathcal{H}_{n})=\mathcal{H}_{n}\otimes \mathcal{H}_{n},
\end{equation}
but in the infinite dimensional case more care is needed, because the
relation $\mathbb{H}=B(\mathcal{H})$ is no more valid. On the contrary,
there are several relevant spaces \cite{FunctAnal1,FunctAnal2}, as the space
of bounded operators $B(\mathcal{H})$ and that of compact operators $C(%
\mathcal{H})$, the space of Hilbert-Schmidt operators $\mathfrak{I}_{2}$ and
that of trace-class operators $\mathfrak{I}_{1}$. Their mutual relations
are:
\begin{equation}
\mathfrak{I}_{1}\subset \mathfrak{I}_{2}\subset C(\mathcal{H})\subset B(%
\mathcal{H}).
\end{equation}
Besides, we recall that $B(\mathcal{H})$ is a Banach space and $C\mathbb{(%
\mathcal{H})}$ a Banach subspace, with the uniform norm $\left\| A\right\|
=\sup_{(\left\| \psi \right\| =1)}\left\| A\psi \right\| ,$ while $%
\mathfrak{I}_{2}$ is a Hilbert space with scalar product $\left\langle
A|B\right\rangle =\mathrm{Tr}\left( A^{\dagger }B\right) $ and norm $\left\|
A\right\| _{2}=$ $\sqrt{\mathrm{Tr}\left( A^{\dagger }A\right) }$ . Finally $%
\mathfrak{I}_{1},$ which is not closed in $B(\mathcal{H})$ with the uniform
norm, is a Banach space with the norm $\left\| A\right\| _{1}=\mathrm{Tr}%
\left( \left| A\right| \right) .$ The following inequalities hold true
\begin{equation}
\left\| A\right\| \leq \left\| A\right\| _{2}\leq \left\| A\right\| _{1}.
\end{equation}

So $\mathfrak{I}_{2}$, the only Hilbert space at our disposal to implement
our definition of tomographic set, is endowed with a topology which, when
restricted to the trace-class operators, is not equivalent to the topology
of $\mathfrak{I}_{1}$. This may have serious consequences. In fact, in the
finite dimensional case, the set $\left\{ P_{\mu }\right\} _{\mu \in M}$ is
complete iff
\begin{equation}
\mathrm{Tr}\left( P_{\mu }A\right) =0\quad \forall \mu \in M\Longrightarrow
A=0.  \label{Azero}
\end{equation}
Such a condition guarantees the full reconstruction of any observable from
its tomograms.

Now, in $\mathfrak{I}_{2}$, Eq. (\ref{Azero}) reads
\begin{equation}
\left\langle P_{\mu }|A\right\rangle =0\quad \forall \mu \in
M\Longrightarrow A=0\quad \&\quad A\in \mathfrak{I}_{2}.
\end{equation}
Then, as $\mathfrak{I}_{2}$ is a $\ast-$ideal in $B(\mathcal{H})$, it may
exists a non-zero operator $B,$ which is bounded but not Hilbert-Schmidt,
such that
\begin{equation}
\mathrm{Tr}\left( P_{\mu }B\right) =0\quad \forall \mu \in M
\end{equation}

In that case, a non-ambiguous reconstruction of two different observables is
impossible when their difference is an operator like $B$. In other words,
different observables may be tomographically separated only when their
difference is Hilbert-Schmidt. For a deeper discussion, see Ref.\cite{Paris}.

Nevertheless there is a second case, when the set $\left\{ P_{\mu }\right\}
_{\mu \in M}$ of trace-class operators is complete even in $\mathfrak{I}_{1}$%
. Then, recalling \cite{FunctAnal3,FunctAnal4} that $\mathfrak{I}_{1}$ is a $%
\ast -$ideal in its dual space $B(\mathcal{H})$:
\begin{equation}
\mathfrak{I}_{1}^{\ast }=B(\mathcal{H}),
\end{equation}
the expression $\mathrm{Tr}\left( P_{\mu }A\right) $ is nothing but the
value of the linear functional $\mathrm{Tr}\left( \cdot A\right) $ in $%
P_{\mu }.$ Hence, Eq.(\ref{Azero}) holds unconditionally
\begin{equation}
\mathrm{Tr}\left( P_{\mu }A\right) =0\quad \forall \mu \in M\Longrightarrow
0=\left\| \mathrm{Tr}\left( \cdot A\right) \right\| =\left\| A\right\|
\Longrightarrow A=0.
\end{equation}
Clearly, this second case is more general: the tomographic map is finer and
is able to better distinguish different observables.

Thus, the finest tomographies are those based on sets of rank-one projectors
which are complete both in $\mathfrak{I}_{2}$ and in $\mathfrak{I}_{1}$. As
a matter of fact, this is the case for the main tomographic sets, as the
photon number and the symplectic tomographic sets.

After the discussion of some of the topological subtleties of the infinite
dimensional case, we are now ready to study an example, which allows for the
construction of a \textit{minimal} tomographic set, i.e., a basis of
rank-one projectors.

\section{Example}

\subsection{A (minimal) tomographic set spanning both $\mathfrak{I}_{2}$ and
$\mathfrak{I}_{1}$}

Let $\{e_{n}\}_{n=1}^{\infty }$ be an orthonormal basis of an Hilbert space $%
\mathcal{H}.$ Now we switch to the Dirac notation, $e_{n}\longleftrightarrow
\left\vert n\right\rangle ,$ and get an orthonormal basis $\{E_{nm}\}=$ $%
\{\left\vert n\right\rangle \left\langle m\right\vert \}_{n,m=1}^{\infty }$
of $\mathfrak{I}_{2}.$ In the basis $\{\left\vert n\right\rangle \},$ we
have
\begin{equation}
\left( E_{nm}\right) _{jk}=\left\langle j\left\vert E_{nm}\right\vert
k\right\rangle =\delta _{jn}\delta _{mk}\,\ ;
\end{equation}
\begin{equation}
\mathrm{Tr}(E_{qp}^{\dagger }E_{nm})=\sum\nolimits_{jk}\left\langle
j\left\vert E_{qp}^{\dagger }\right\vert k\right\rangle \left\langle
k\left\vert E_{nm}\right\vert j\right\rangle =\sum\nolimits_{jk}\delta
_{jp}\delta _{qk}\delta _{kn}\delta _{mj}=\delta _{qn}\delta _{pm}\,\ .
\end{equation}
A Hermitian orthogonal basis may be constructed with the compact operators
\begin{equation}
E_{nm}^{+}=\frac{1}{2}\left( E_{nm}+E_{nm}^{\dagger }\right) \quad \left(
n\leq m\right) ;\quad E_{nm}^{-}=\frac{i}{2}(E_{nm}-E_{nm}^{\dagger })\quad
\left( n>m\right) .
\end{equation}
The Hermitian basis is readily diagonalizable: for $n\neq m$ the set $%
\left\{ E_{nm}^{+},E_{nm}^{-}\right\} _{n, m}$ is isospectral, with simple
eigenvalues $\pm 1/2$ and respective eigenvectors
\begin{equation}
\left| \Psi _{nm}^{+,\pm }\right\rangle =\frac{1}{\sqrt{2}}\left( \left|
n\right\rangle \pm \left| m\right\rangle \right) \quad ;\quad \left| \Psi
_{nm}^{-,\pm }\right\rangle =\frac{1}{\sqrt{2}}\left( \left| m\right\rangle
\pm i\left| n\right\rangle \right) ,
\end{equation}
where $\pm$ label the eigenvalues. Their associated projectors
\begin{equation}
P_{nm}^{+ ,\pm }=\left| \Psi _{nm}^{+ ,\pm }\right\rangle \left\langle \Psi
_{nm}^{+,\pm }\right| ,\quad P_{nm}^{-,\pm }=\left| \Psi _{nm}^{-,\pm
}\right\rangle \left\langle \Psi _{nm}^{- ,\pm }\right|,
\end{equation}
together with the diagonal $(n=m)$ projectors
\begin{equation}
P_{nn}=\left| n\right\rangle \left\langle n\right| ,
\end{equation}
are a tomographic set. In fact, as
\begin{equation}
P_{nm}^{+,\pm }=\frac{1}{2}(P_{nn}+P_{mm})\pm E_{nm}^{+}\quad ,\quad
P_{nm}^{-,\pm }=\frac{1}{2}(P_{nn}+P_{mm})\pm E_{nm}^{-},
\end{equation}
the set contains a basis of $\mathfrak{I}_{2}$ of rank-one projectors.
Moreover, the set is complete in $\mathfrak{I}_{1}.$ In fact, assume that
the linear functional $\mathrm{Tr}(A\cdot ),$ with $A\in B(\mathcal{H}),$
vanishes on the tomographic set. Then
\begin{equation}
\mathrm{Tr}(AP_{nn})=\left\langle n\left| A\right| n\right\rangle =0\quad
\forall n,
\end{equation}
so that the diagonal matrix elements of $A$ are zero. Bearing this in mind,
we have
\begin{equation*}
\mathrm{Tr}(AP_{nm}^{+,\pm })=\frac{1}{2}\mathrm{Tr}(A\left(
P_{nn}+P_{mm}\pm 2E_{nm}^{+}\right) )=\pm \frac{1}{2}(\left\langle m\left|
A\right| n\right\rangle +\left\langle n\left| A\right| m\right\rangle )=0,
\end{equation*}
\begin{equation*}
\mathrm{Tr}(AP_{nm}^{-,\pm })=\frac{1}{2}\mathrm{Tr}(A\left(
P_{nn}+P_{mm}\pm 2E_{nm}^{-}\right) )=\pm \frac{i}{2}(\left\langle m\left|
A\right| n\right\rangle -\left\langle n\left| A\right| m\right\rangle )=0,
\end{equation*}
which yield
\begin{equation}
\left\langle m\left| A\right| n\right\rangle =0\quad \forall
m,n\Leftrightarrow A=0,
\end{equation}
so that $A$ is the zero operator.

Finally we observe that a \textsl{minimal} tomographic set, i.e. a basis of
rank-one projectors, may be chosen by taking just one projector from each
pair $P_{nm}^{+,\pm }$ with $n<m$, only one projector from each pair $%
P_{nm}^{-,\pm }$ with $n>m$ and all the diagonal $P_{nn}$'s. Such a minimal
set is obviously complete both in $\mathfrak{I}_{2}$ and in $\mathfrak{I}%
_{1}.$

\subsection{The corresponding resolution of the unity}

We now evaluate explicitly the resolution of unity determined by the full
(non-minimal) set of projectors. To do this, we start from the
representation of a (bounded) operator $B$ as
\begin{equation}
B=\sum_{n,m}\left\langle n\left| B\right| m\right\rangle \left|
n\right\rangle \left\langle m\right| .
\end{equation}
In view of the decomposition of any operator as a sum of two selfadjoint
operators
\begin{equation}
B=\frac{1}{2}(B+B^{\dagger })-i(\frac{i}{2}(B-B^{\dagger })),
\end{equation}
we may assume $B$ selfadjoint. Then the identity holds:
\begin{eqnarray}
\left\langle n\left| B\right| m\right\rangle &=&\frac{1}{2}\left[
\left\langle\Psi _{nm}^{+,+ }\left| B\right| \Psi _{nm}^{+,+ }\right\rangle
-\left\langle \Psi _{nm}^{+,- }\left| B\right| \Psi _{nm}^{+,-
}\right\rangle \right]  \notag \\
&+&\frac{i}{2}\left[\left\langle \Psi _{nm}^{-,+ }\left| B\right| \Psi
_{nm}^{-,+ }\right\rangle -\left\langle \Psi _{nm}^{-,- }\left| B\right|
\Psi _{nm}^{-,- }\right\rangle \right].
\end{eqnarray}
In other terms:
\begin{eqnarray}
\left\langle n\left| B\right| m\right\rangle =\frac{1}{2}\left[ \mathrm{Tr}%
(BP_{nm}^{+,+})-\mathrm{Tr}(BP_{nm}^{+,-})+i (\mathrm{Tr}(BP_{nm}^{-,+})-%
\mathrm{Tr}(BP_{nm}^{-,-})\right].  \notag
\end{eqnarray}
Thus, we get the reconstruction formula
\begin{equation*}
B=\sum_{n,m}\frac{1}{2}\left| n\right\rangle \left\langle m\right| \left[
\mathrm{Tr}(BP_{nm}^{+,+})-\mathrm{Tr}(BP_{nm}^{+,-})+i\mathrm{Tr}%
(BP_{nm}^{-,+})-i\mathrm{Tr}(BP_{nm}^{-,-})\right],
\end{equation*}
or, equivalently,
\begin{eqnarray}
B &=&\sum_{n}P_{nn}\mathrm{Tr}(BP_{nn})+\sum_{n<m}E_{nm}^{+}\left[ \mathrm{Tr%
}(BP_{nm}^{+,+})-\mathrm{Tr}(BP_{nm}^{+,-})\right]  \notag \\
&+&\sum_{n<m}E_{nm}^{-}\left[ \mathrm{Tr}(BP_{nm}^{-,+})-\mathrm{Tr}%
(BP_{nm}^{-,-})\right] .
\end{eqnarray}
Upon introducing a third label $\alpha $ to enumerate the $P^{\pm,\pm}$'s,
we obtain the resolution of the unity as
\begin{equation}
\hat{\mathbb{I}}=\sum_{n}\left| P_{nn}\right\rangle \left\langle
P_{nn}\right| +\sum_{n<m,\alpha }\left| K_{nm}^{\alpha }\right\rangle
\left\langle P_{nm}^{\alpha }\right| \,\ ,
\end{equation}
where
\begin{equation}
\left| K_{nm}^{+,\pm }\right\rangle =\pm E_{nm}^{+},\quad \left|
K_{nm}^{-,\pm }\right\rangle =\pm E_{nm}^{-} \,\ .
\end{equation}

\section{Families of operators generating tomographic sets}

An interesting question is how to construct tomographic sets. We will answer
this question by considering how some of the main tomographic sets are
generated, so providing also a few well known examples to discuss.

We start with a fiducial Hermitian operator $\hat{T}_{0}$ and act on it with
a family of unitary operators $\left\{ U_{\mu }\right\} $, depending on some
parameters $\mu \in M,$ to generate a family of (iso-spectral)$\ $Hermitian
operators
\begin{equation}
\hat{T}_{\mu }=U_{\mu }\hat{T}_{0}U_{\mu }^{\dagger }.
\end{equation}
Assuming $\hat{T}_{0}$ to be generic, i.e. with simple eigenvalues, the
action of $U_{\mu }$ on the rank-one projectors associated with the
eigenstates $\left\{ \left\vert \psi _{n}^{0}\right\rangle \right\} _{n\in
\mathbb{N}}$ of $\hat{T}_{0}\ $gives rise to a set of projectors,
corresponding to the eigenstates $\left\{ \left\vert \psi _{\mu
,n}\right\rangle \right\} _{n}=$ $\left\{ U_{\mu }\left\vert \psi
_{n}^{0}\right\rangle \right\} _{n}$of $\hat{T}_{\mu }$
\begin{equation}
P_{\mu ,n}=U_{\mu }P_{n}^{0}U_{\mu }^{\dagger },\quad \mu \in M.
\label{Pgen}
\end{equation}
We observe that Eq. (\ref{Pgen}) suggests one could start with a fiducial
rank-one projector $P_{0}$ as operator $\hat{T}_{0}$ and act on it with the
unitary family to generate a set of projectors. However the use of a generic
operator $\hat{T}_{0}$ allows, if the set $\left\{ P_{\mu ,n}\right\} ,$ $%
\left( \mu ,n\right) \in M\times \mathbb{N},$ is tomographic, to obtain at
once that the tomograms of any density operator $\hat{\rho}$ satisfy
\begin{equation}
\sum_{n}\mathcal{T}_{\rho }(\mu ,n)=\sum_{n}\mathrm{Tr}\left( \hat{\rho}%
P_{\mu ,n}\right) =\sum_{n}\left\langle \psi _{\mu ,n}\left\vert \hat{\rho}%
\right\vert \psi _{\mu ,n}\right\rangle =1,\quad \forall \mu \in M.
\end{equation}

Such an identity is of capital importance, because it allows for the
probabilistic interpretation of the tomographic map $\mathcal{T}$, as for
any given $\mu$ the tomogram $\mathcal{T}_{\rho }(\mu)$ is a marginal
probability distribution.

Then the question is, how the operator $\hat{T}_{0}$ and the unitary family $%
\left\{ U_{\mu }\right\} $ have to be chosen to generate a tomographic set $%
\left\{ P_{\mu ,n}\right\} ?$ Or more simply, when is the set $\left\{
P_{\mu ,n}\right\} $ tomographic?

We may preliminarily state a negative answer.

\noindent\textbf{Proposition.} \textit{The set $\left\{ P_{\mu
,n}\right\}
$ is not tomographic, if it exists a decomposition $\mathcal{H}=%
\mathcal{H}_{1}\oplus \mathcal{H}_{2}$ invariant under the action
of both the operator $\hat{T}_{0}$ and the unitary family $\left\{
U_{\mu }\right \} $.}

\noindent \textbf{Proof.} In fact, the set
$\left\{ P_{\mu ,n}\right\} $ is not complete, as the non-zero
operator $\left\vert \varphi _{1}\right\rangle \left\langle
\varphi _{2}\right\vert $, with $\left\vert \varphi
_{1}\right\rangle \in \mathcal{H}_{1}$ and $\left\vert \varphi
_{2}\right\rangle \in \mathcal{H}_{2},$ is orthogonal to the whole set $%
\left\{ P_{\mu ,n}\right\} :$%
\begin{equation}
\mathrm{Tr}(\left\vert \varphi _{1}\right\rangle \left\langle \varphi
_{2}\right\vert P_{\mu ,n})=\mathrm{Tr}(\left\vert \varphi _{1}\right\rangle
\left\langle \varphi _{2}\right\vert U_{\mu }\left\vert \psi
_{n}^{0}\right\rangle \left\langle \psi _{n}^{0}\right\vert U_{\mu
}^{\dagger })=0\quad \forall \mu ,n
\end{equation}
because $\left\langle \varphi _{2}\right\vert U_{\mu }\left\vert \psi
_{n}^{0}\right\rangle =0$ or $\left\langle \psi _{n}^{0}\right\vert U_{\mu
}^{\dagger }\left\vert \varphi _{1}\right\rangle =0,$ according to the case $%
\left\vert \psi _{n}^{0}\right\rangle \in \mathcal{H}_{1}$ or $\left\vert
\psi _{n}^{0}\right\rangle \in \mathcal{H}_{2}. \,\ \blacksquare $

\noindent\textbf{Example: the squeeze \textquotedblleft
tomography\textquotedblright . }It is generated by the family of
(iso-spectral)$\ $Hermitian operators depending on two real parameters
\begin{equation}
\hat{T}_{sq}(\mu ,\nu )=S(\mu ,\nu )\hat{a}^{\dagger }\hat{a}S^{\dagger
}(\mu ,\nu ),\quad \mu ,\nu \in \mathbb{R},  \label{Squeezetom}
\end{equation}
where the unitary operators $\left\{ S(\mu ,\nu )\right\} $ depend
quadratically on the harmonic oscillator creation and annihilation operators
$\hat{a}^{\dagger },\hat{a},$ see \emph{Ref(Squeeze).} Then, both the
fiducial operator $\hat{a}^{\dagger }\hat{a}$ and the unitary family $%
\left\{ S(\mu ,\nu )\right\} $ commute with the Parity operator. So, the
squeeze \textquotedblleft tomography\textquotedblright\ is not a true
tomography. Nevertheless, we can get a true tomographic set by a restriction
to the subspace of even wave functions. Only there the existence of an
inversion formula is granted. $\Box$

This example shows that the answer depends on a joint property of $\hat{T}%
_{0}\ $and $\left\{ U_{\mu }\right\} ,$ i.e. the relation between the
commutants $\hat{T}_{0}^{\prime }$ of the fiducial operator and $\left\{
U_{\mu }\right\} ^{\prime }$ of the unitary family.

So, we may restate the previous proposition as a \textit{necessary
condition}:

 \noindent \textbf{Proposition.} \textit{If the set
$\left\{ P_{\mu
,n}\right\} $ is tomographic, then the family $\{\hat{T}%
_{0},\left\{ U_{\mu }\right\} \}$ is irreducible or, equivalently,
the intersection of the commutants $\hat{T}_{0}^{\prime }$ \textit{and }$%
\left\{ U_{\mu }\right\} ^{\prime }$ is trivial: $\hat{T}%
_{0}^{\prime }\cap \left\{ U_{\mu }\right\} ^{\prime }=\left\{
1\right\} .$}

For instance, by changing the unitary family $\left\{ S(\mu ,\nu )\right\} $
or the starting operator $\hat{a}^{\dagger }\hat{a}$ we may obtain from Eq.(%
\ref{Squeezetom}) tomographic families of selfadjoint operators, as in the
following

\noindent \textbf{Example: the symplectic tomography.} It is generated by
the same family of unitary operators $\left\{ S(\mu ,\nu )\right\} $ of the
squeeze ``tomography'', with the position operator $\hat{Q}$ as fiducial
operator:
\begin{equation}
\hat{T}(\mu ,\nu )=S(\mu ,\nu )\hat{Q}S^{\dagger }(\mu ,\nu )=\mu \hat{Q}%
+\nu \hat{P},\quad \mu ,\nu \in \mathbb{R}
\end{equation}
where $\hat{P}$ is the momentum operator. The spectrum is continuous. The
(improper) eigenvectors $\left\{ \left| X\mu \nu \right\rangle \right\} $ of
$\hat{T}(\mu ,\nu )$ stem out from those of the position: $\hat{Q}\left|
X\right\rangle =X\left| X\right\rangle ,X\in \mathbb{R}$. In the position
representation $\{\left| q\right\rangle \}$, for $\nu \neq 0$:
\begin{equation}
\left\langle q|X\mu \nu \right\rangle =\left\langle q\right| S(\mu ,\nu
)\left| X\right\rangle =\frac{1}{\sqrt{2\pi |\nu |}}\exp \left[ -i(\frac{\mu
}{2\nu }q^{2}-\frac{X}{\nu }q)\right] ,
\end{equation}
with
\begin{equation}
\left\langle X^{\prime }\mu \nu |X\mu \nu \right\rangle =\delta \left(
X-X^{\prime }\right)
\end{equation}

Then the resolution of unity in matrix form reads \cite{PLA2}:
\begin{equation*}
\int \frac{dX}{2\pi }d\mu d\nu \left\langle y|\exp \left[ i\left( X-\mu \hat{%
Q}-\nu \hat{P}\right) \right] |y^{\prime }\right\rangle \left\langle
q^{\prime }\left\vert X\mu \nu \right\rangle \left\langle X\mu \nu
\right\vert q\right\rangle =\delta (q-y)\delta (q^{\prime }-y^{\prime }).\Box
\notag
\end{equation*}

The irreducibility condition of $\{\hat{T}_{0},\left\{ U_{\mu }\right\} \}$
is too poor to get a sufficient condition. For this, more hypotheses must be
added. For instance, the family of unitary operators $\left\{ U_{\mu
}\right\} $ may be chosen as a representation of a group $G:$ $\mu
\leftrightarrow g\in G.$ When this representation is analytic in some
neighborhood of $\mu =0,$ then the set $\left\{ P_{\mu ,n}\right\} $ is
tomographic. This is the case of the coherent state and of the photon number
tomographic sets, discussed in the next section. The analyticity condition
may be be substituted by other weaker hypotheses, but these further
conditions are needed, as the following finite dimensional counter-example
shows.

\noindent \textbf{Example.} On $\mathcal{H}=\mathbb{C}^{2}$, take for $%
\left\{ U_{\mu }\right\} $ the family
\begin{equation}
U_{1}=\hat{I}=\left[
\begin{array}{cc}
1 & 0 \\
0 & 1
\end{array}
\right] ,\quad U_{-1}=\hat{P}=\left[
\begin{array}{cc}
1 & 0 \\
0 & -1
\end{array}
\right] ,
\end{equation}
which represents the group $Z_{2}.$ As fiducial operator choose
\begin{equation}
\hat{T}_{0}=\left[
\begin{array}{cc}
\alpha & \beta \\
\beta ^{\ast } & \gamma
\end{array}
\right] ,\quad \alpha ,\gamma \in \mathbb{R},\quad \beta \neq 0,
\end{equation}
whose eigenvalues are $\lambda =[\alpha +\gamma \pm \sqrt{\left( \alpha
-\gamma \right) ^{2}+4\left\vert \beta \right\vert ^{2}}]/2.$ The condition $%
\beta \neq 0$ implies that $\hat{T}_{0}^{\prime }\cap \left\{ U_{\mu
}\right\} ^{\prime }=\left\{ 1\right\} ,$ so that $\{\hat{T}_{0},\left\{
U_{\mu }\right\} \}$ is irreducible. The isospectral family is
\begin{equation}
\hat{T}_{0}=\left[
\begin{array}{cc}
\alpha & \beta \\
\beta ^{\ast } & \gamma
\end{array}
\right] ,\quad \hat{P}\hat{T}_{0}\hat{P}=\left[
\begin{array}{cc}
\alpha & -\beta \\
-\beta ^{\ast } & \gamma
\end{array}
\right] .
\end{equation}
Then, if $\beta $ is real and $\alpha =\gamma =0$, $\hat{P}\hat{T}_{0}\hat{P}%
=-\hat{T}_{0}$ and the family $\left\{ U_{\mu }\right\} $ does not displace $%
\hat{T}_{0}.$ Otherwise, the isospectral family has two different operators.
But from Ref. \cite{PLA2} we know that three different operators are needed
to get a basis of rank-one projectors. In any case, the set of the
projectors associated with the eigenvectors of the isospectral family is
\textit{not} tomographic. $\Box $

This simple example shows that, starting from $\hat{T}_{0},$ the unitary
family has to generate a number of different isospectral operators $\hat{T}%
_{\mu }$ sufficient to get a complete set of rank-one (eigen-) projectors $%
\{P_{\mu }\}$. So, the strong condition of analyticity is only a suitable
way to obtain such a complete set. However, the unitary family needs not to
be a representation of any group, as the case of the countable tomographic
set of section 4 shows:

\noindent \textbf{Example.} Take as fiducial operator
\begin{equation}
\hat{T}_{0}=\mathrm{diag}\left[ 1,-1,0,...,0,...\right] .  \label{Texamp}
\end{equation}
The first $2\times 2$ block is $\sigma _{3},$ one of the Pauli matrices:
\begin{equation}
\sigma _{1}=\left[
\begin{array}{cc}
0 & 1 \\
1 & 0
\end{array}
\right] ,\quad \sigma _{2}=\left[
\begin{array}{cc}
0 & -i \\
i & 0
\end{array}
\right] ,\quad \sigma _{3}=\left[
\begin{array}{cc}
1 & 0 \\
0 & -1
\end{array}
\right]
\end{equation}
By means of the rules
\begin{equation}
\left[ \sigma _{j},\sigma _{k}\right] _{+}=2\delta _{jk};\quad \left[ \sigma
_{j},\sigma _{k}\right] =2i\varepsilon _{jkl}\sigma _{l}
\end{equation}
we get
\begin{equation}
\exp \left[ -i\vec{\sigma}\cdot \vec{n}\frac{\phi }{2}\right] =\cos \frac{%
\phi }{2}-i\vec{\sigma}\cdot \vec{n}\sin \frac{\phi }{2}
\end{equation}
so that
\begin{equation}
\exp \left[ -i\vec{\sigma}\cdot \vec{n}_{1}\frac{\pi }{2}\right] \sigma
_{3}\exp \left[ i\vec{\sigma}\cdot \vec{n}_{1}\frac{\pi }{2}\right] =\sigma
_{1},\quad \vec{n}_{1}=\frac{1}{\sqrt{2}}\left( 1,0,1\right) ,
\end{equation}
and
\begin{equation}
\exp \left[ -i\vec{\sigma}\cdot \vec{n}_{2}\frac{\pi }{4}\right] \sigma
_{1}\exp \left[ i\vec{\sigma}\cdot \vec{n}_{2}\frac{\pi }{4}\right] =\sigma
_{2},\quad \vec{n}_{2}=\left( 0,0,1\right)
\end{equation}
We may define the unitary operators $S(\vec{n},\frac{\phi }{2})$ by
embedding $\exp \left[ -i\vec{\sigma}\cdot \vec{n}\frac{\phi }{2}\right] $
as first $2\times 2$ block into a zero matrix.

Then, consider the unitary selfadjoint commuting operators $U_{1,n}$ and $%
U_{2,m}$ which interchange the vector components $1,n$ and $2,m$
respectively. Upon multiplying $S$'s and $U$'s operators, we construct a
family (not a group) of unitary operators which displace the operator $\hat{T%
}_{0}$ of Eq.(\ref{Texamp}) and generate the tomographic family of compact
operators $\{E_{n,m}^{+},E_{n,m}^{-}\}$ of section 4:
\begin{eqnarray}
&&U_{1,n}U_{2,m}S(\vec{n}_{1},\frac{\pi }{2})\hat{T}_{0}S^{\dag }(\vec{n}%
_{1},\frac{\pi }{2})U_{2,m}U_{1,n}=2E_{n,m}^{+}\,\ , \\
&&U_{1,n}U_{2,m}S(\vec{n}_{2},\frac{\pi }{4})S(\vec{n}_{1},\frac{\pi }{2})%
\hat{T}_{0}S^{\dag }(\vec{n}_{1},\frac{\pi }{2})S^{\dag }(\vec{n}_{2},\frac{%
\pi }{4})U_{2,m}U_{1,n}=2E_{n,m}^{-}\,\ . \,\ \Box  \notag
\end{eqnarray}

These last two examples show that neither the hypothesis of irreducibility
nor the condition of analyticity of the representation $\left\{ U_{\mu
}\right\} $ of the group are necessary. However, the analytic dependence on
the parameter $\mu $ together with the irreducibility of $\left\{ \hat{T}%
_{0},\left\{ U_{\mu }\right\} \right\} $ are sufficient for constructing a
tomographic set, as it is elucidated in the next section.

\section{Skewness}

From a geometrical point of view, tomographic sets are sets of
\textquotedblleft skew\textquotedblright\ projectors. In other words, the
\textit{skewness} of the projectors' set denotes its completeness from a
geometrical point of view. For instance, in the qu-bit case of the spin
tomography, the manifold of rank-one projectors in the real space $\mathbb{R}%
^{4}$ of Hermitian operators is the Bloch sphere $S^{2}:\left\{ \ \left(
x^{2}\right) ^{2}+\left( x^{3}\right) ^{2}+\left( x^{4}\right)
^{2}=1/4\right\} $ placed in the plane $x^{1}=1/2$. Then any set of four
points of $S^{2},$ not lying on the equator, is skew as the corresponding
projectors generate the whole space \cite{PLA2}. Then we give the following:

\noindent \textbf{Definition.} A set of projectors is globally
skew when it spans the whole Hilbert space.

Thus, any tomographic set is \textit{globally skew} as it is
complete. Besides:

\noindent \textbf{Definition.} A set $\gamma $ of projectors
containing $P_{0}$ is locally skew in $P_{0}$ if any neighborhood
of $P_{0}$ contains a skew subset of $\gamma .$

Back to the qu-bit case, any set of points on $S^{2},$ not lying on the
equator and with a limit point $P_{0},$ is locally skew\textit{\ }in $P_{0}.$
For the infinite dimensional case, we observe that the countable tomographic
set of section 4 is skew globally but not locally. Perhaps the simplest case
of a tomographic set which is skew globally and locally is provided by the
following

\noindent \textbf{Example: the coherent state tomography.} This tomographic
set, which is studied in Ref. \cite{PLA3}, is generated by the displacement
operators $\left\{ \mathcal{D}\left( \alpha \right) \right\} $ depending on
a complex parameter $\alpha $%
\begin{equation}
\mathcal{D}\left( \alpha \right) =\exp \left( \alpha \hat{a}^{\dagger
}-\alpha ^{\ast }\hat{a}\right) ,\quad \alpha \in \mathbb{C},
\end{equation}
which acting on the projector $\left\vert 0\right\rangle \left\langle
0\right\vert $ of the vacuum Fock state, $\hat{a}\left\vert 0\right\rangle
=0,$ yield the projectors
\begin{equation}
\left\vert \alpha \right\rangle \left\langle \alpha \right\vert =\mathcal{D}%
\left( \alpha \right) \left\vert 0\right\rangle \left\langle 0\right\vert
\mathcal{D}\left( \alpha \right) ^{\dagger },\quad \alpha \in \mathbb{C},
\end{equation}
associated to the usual coherent states
\begin{equation}
\left\vert \alpha \right\rangle =\exp (-\frac{\left\vert \alpha \right\vert
^{2}}{2})\exp \left( \alpha \hat{a}^{\dagger }\right) \exp \left( -\alpha
^{\ast }\hat{a}\right) \left\vert 0\right\rangle =\exp (-\frac{\left\vert
\alpha \right\vert ^{2}}{2})\sum_{j=0}^{\infty }\frac{\alpha ^{j}}{n!}\hat{a}%
^{\dagger j}\left\vert 0\right\rangle \,\ .
\end{equation}

We recall that the coherent states are a (over-) complete set in the Hilbert
space $\mathcal{H}.$ Any bounded set containing a limit point $\alpha _{0}$
in the complex $\alpha -$plane defines a complete set of coherent states
containing a limit point, the coherent state $\left\vert \alpha
_{0}\right\rangle ,$ in the Hilbert space $\mathcal{H}$. In particular, any
Cauchy sequence $\{\alpha _{k}\}$ of complex numbers defines a Cauchy
sequence of coherent states $\left\{ \left\vert \alpha _{k}\right\rangle
\right\} ,$ which is a complete set. The same holds for any extracted
subsequence. This completeness property holds as $\exp \left( \left\vert
\alpha \right\vert ^{2}/2\right) \left\langle \alpha |\psi \right\rangle $
is an entire analytic function of the complex variable $\alpha ^{\ast },$
for any $\left\vert \psi \right\rangle \in \mathcal{H},$ with a non-isolated
zero in $\alpha _{0}^{\ast }$. Then
\begin{equation}
\left\langle \alpha _{k}|\psi \right\rangle =0\quad \forall k\Rightarrow
\left\vert \psi \right\rangle =0.
\end{equation}

Besides, any bounded operator $A$ may be completely reconstructed from its
diagonal matrix elements $\left\langle \alpha _{k}\left\vert A\right\vert
\alpha _{k}\right\rangle .$ In fact, $\exp ( \left\vert \alpha \right\vert
^{2}/2+\left\vert \beta \right\vert ^{2}/2) \left\langle \alpha \left\vert
A\right\vert \beta \right\rangle $ is an analytical function of the complex
variables $\alpha ^{\ast },\beta ,$\ so it is uniquely determined by its
value $\exp( \left\vert \alpha \right\vert ^{2}) \left\langle \alpha
\left\vert A\right\vert \alpha \right\rangle $ on the diagonal $\beta
=\alpha .$ This is an entire function of the real variables $\Re \alpha ,\Im
\alpha ,$ which is in turn uniquely determined by its values on any set with
an accumulation point.

The rank-one projectors associated to a complete set of coherent states are
complete in the Hilbert space $\mathfrak{I}_{2}$. In particular, any Cauchy
sequence $\left\{ \left\vert \alpha _{k}\right\rangle \right\} $ generates a
tomographic set $\left\{ \left\vert \alpha _{k}\right\rangle \left\langle
\alpha _{k}\right\vert \right\} $. In fact, bearing in mind the previous
remark on the reconstruction of a bounded operator, it results
\begin{equation}
\mathrm{Tr}(A\left\vert \alpha _{k}\right\rangle \left\langle \alpha
_{k}\right\vert )=\left\langle \alpha _{k}\left\vert A\right\vert \alpha
_{k}\right\rangle =0\quad \forall k\Rightarrow A=0\quad \&\quad A\in B(%
\mathcal{H}).
\end{equation}
This shows that a tomographic set of coherent state projectors is complete
even in $\mathfrak{I}_{1}.$ So it is globally skew. Moreover, any extracted
subsequence $\left\{ \left\vert \alpha _{k_{j}}\right\rangle \left\langle
\alpha _{k_{j}}\right\vert \right\} $ is again complete, so $\left\{
\left\vert \alpha _{k}\right\rangle \left\langle \alpha _{k}\right\vert
\right\} $ is locally skew in its limit point. The case when $\alpha $
varies in the whole complex plane and the associated reconstruction formula
are discussed in \cite{PLA3}. $\Box $

The same considerations hold for the following example, strictly connected
the previous one.

\noindent \textbf{Example: the photon number tomography.} It is generated by
the irreducible family $\{\hat{a}^{\dagger }\hat{a},\left\{ \mathcal{D}%
\left( \alpha \right) \right\} \},$ where the displacement operators $%
\left\{ \mathcal{D}\left( \alpha \right) \right\} $ act on the same fiducial
operator $\hat{a}^{\dagger }\hat{a}$ of the squeeze \textquotedblleft
tomography\textquotedblright ,
\begin{equation}
\hat{T}(\alpha )=\mathcal{D}\left( \alpha \right) \hat{a}^{\dagger }\hat{a}%
\mathcal{D}\left( \alpha \right) ^{\dagger },\quad \alpha \in \mathbb{C}
\end{equation}
The family of selfadjoint operators $\hat{T}(\alpha )$ has the spectrum of
the number operator $\hat{a}^{\dagger }\hat{a},$ and eigenvectors $%
\left\vert n\alpha \right\rangle =\mathcal{D}\left( \alpha \right)
\left\vert n\right\rangle .$ With $\alpha =\left( \nu +i\mu \right) /\sqrt{2}%
,$ in the position representation $\left\langle y|n\alpha \right\rangle $ is

\begin{eqnarray}
\int dq\left\langle y|\mathcal{D}\left( \alpha \right) |q\right\rangle
\left\langle q|n\right\rangle &=&\int dq\delta \left( y-q-\nu \right) \exp %
\left[ i\left( \mu q+\mu \nu /2\right) \right] \left\langle q|n\right\rangle
\notag \\
&=&\exp \left[ i\left( \mu y-\mu \nu /2\right) \right] \left\langle y-\nu
|n\right\rangle ,  \label{alfaenne}
\end{eqnarray}
where the $n$-th Hermite function $\left\langle q|n\right\rangle $ is
\begin{equation}
\left\langle q|n\right\rangle =(\sqrt{\pi }2^{n}n!)^{-1/2}\exp (-\frac{1}{2}%
q^{2})H_{n}(q).  \label{accaenne}
\end{equation}

We recall that the photon number projectors' set, containing the complete
set of the coherent state projectors, is a tomographic set complete both in $%
\mathfrak{I}_{2}$ and $\mathfrak{I}_{1}.$ For the same reason, any Cauchy
sequence $\left\{ \left\vert n\alpha _{k}\right\rangle \left\langle n\alpha
_{k}\right\vert \right\} $ is locally skew in its limit point.

The whole set of photon number projectors generates the resolution of unity
\begin{equation}
\mathbb{I}=\sum\limits_{n=0}^{\infty }\int \frac{d^{2}\alpha }{\pi }%
K^{\left( s\right) }\left( n,\alpha \right) \mathrm{Tr}(\left\vert n\alpha
\right\rangle \left\langle n\alpha \right\vert \cdot )  \label{Photoris}
\end{equation}
The Gram-Schmidt operator $K^{\left( s\right) }$ is given by
\begin{equation}
K^{\left( s\right) }\left( n,\alpha \right) =\frac{4}{s^{2}-1}\left( \frac{%
s+1}{s-1}\right) ^{n}\mathcal{D}\left( \alpha \right) \left( \frac{s-1}{s+1}%
\right) ^{\hat{a}^{\dagger }\hat{a}}\mathcal{D}^{\dagger }\left( \alpha
\right) .
\end{equation}
Here $s$ is a real parameter, $-1<s<1,$ which labels the family of
equivalent kernels $K^{\left( s\right) }\left( n,\alpha \right) .$ This
formula corrects the corresponding expressions given in \cite{PLA2}.

The check of the matrix form of the resolution of the unity, Eq.(\ref
{Photoris}), in the position representation is done in \cite{PLA3} and
yields:
\begin{equation}
\sum\limits_{n=0}^{\infty }\int \frac{d^{2}\alpha }{\pi }\left\langle
y^{\prime }\left\vert n\alpha \right\rangle \left\langle n\alpha \right\vert
x^{\prime }\right\rangle \left\langle x|K^{\left( s\right) }\left( n,\alpha
\right) |y\right\rangle =\delta (x-x^{\prime })\delta (y-y^{\prime }),
\end{equation}
for any allowed $s,$ as it was expected. $\Box$

\section{Conclusions}

To conclude we summarize the main points of the paper. We have reviewed the
tomographic methods to map the vectors (and non-negative Hermitian
trace-class operators) in abstract Hilbert spaces onto standard probability
distributions and established conditions for the existence of the inverse
transform both for the finite and infinite-dimensional cases.

In the infinite-dimensional case all the known examples of tomographies,
like symplectic tomography, coherent state tomography, photon number
tomography, squeeze tomography, were considered in the suggested framework
of the existence of tomographic sets as over-complete bases of rank-one
projectors. Any such a basis determines a completeness relation, that is a
resolution of the (super-) identity operator, acting on the space of the
bounded operators on the initial Hilbert space $\mathcal{H},$ which is
expressed generally in terms of the rank-one projectors and the
corresponding Gram-Schmidt operators.

\noindent \textbf{Acknowledgements} Vladimir Man'ko thanks University
``Federico II''\ and INFN, Sezione di Napoli, and the Organizers of the
Palermo conference ``TQMFA 2005''\ for the hospitality and support.

\end{document}